# Diffusion-based Generative Machine Learning Model for Predicting Crack Propagation in Aluminum Nitride at the Atomic Scale

Jiali Lu, Shengfeng Yang*

Purdue University

**Abstract**

Predicting atomic-scale crack propagation in aluminum nitride (AlN) is critical for semiconductor reliability but remains prohibitively expensive via molecular dynamics (MD). We develop a diffusion-based generative machine learning model to predict atomic-scale crack propagation in AlN, a critical semiconductor material, by conditioning solely on initial microstructure embeddings. Trained on MD simulations of single-crack systems, the model achieves a significant speedup while accurately forecasting dynamic fracture processes, including stress-driven crack initiation, crack branching, and atomic-scale bridging ligaments. Crucially, it demonstrates inherent physical fidelity by reproducing material-intrinsic mechanisms while disregarding periodic boundary artifacts, and generalizes to unseen multi-crack configurations. Validation against MD ground truth confirms the model's capability to capture complex fracture physics without auxiliary stress or energy data, enabling rapid exploration of crack-mediated failure for semiconductor reliability optimization.

## 1. Introduction

Aluminum nitride (AlN) is a critical semiconductor material due to its exceptional thermal conductivity, electrical insulation, mechanical strength, and resistance to radiation and corrosion [1]. These properties make it indispensable for demanding applications in power electronics, high-frequency devices (including 5G), electric vehicles, and aerospace systems [1]. Heteroepitaxial growth on substrates is a key fabrication technique for AlN films, offering precise control over layer thickness and enabling complex geometries [2-4]. However, a significant challenge persists: the formation of cracks during the cooling phase of fabrication and operation [5-9]. These cracks compromise device integrity, reliability, and yield, driving up manufacturing costs.

*Corresponding author
Email address: shengfengyang@purdue.edu (Shengfeng Yang)

Predicting crack initiation and propagation in semiconductors like AlN is inherently complex, requiring an understanding of dynamic atomic-scale mechanisms, especially as device features shrink towards the nanoscale. Traditional continuum-level methods, such as the Finite Element Method (FEM), lack the resolution to capture these atomistic details. While Molecular Dynamics (MD) simulations provide the necessary fidelity, their extreme computational cost severely limits their utility for exploring complex crack evolution over relevant timescales or across diverse microstructural conditions.

Machine learning (ML) offers a promising pathway to overcome these computational barriers [10, 11]. ML models [12-14] have demonstrated success in predicting various material properties and behaviors, including grain boundary energy [15, 16], segregation energy [17] and even full-field quantities like stress distribution [18-26] and fracture behavior [27-30]. Deep generative models, in particular, show significant potential for capturing complex spatial patterns. Conditional Generative Adversarial Networks (cGANs) have been applied to predict stress field [18, 25] and grain boundary evolution [31, 32]. and grain boundary evolution. More recently, diffusion-based generative models have emerged as powerful tools for high-fidelity image generation, including applications in predicting fracture in simplified materials modeled with Leonard-Jones potentials [33-35].

However, a critical gap remains: the application of advanced generative ML models, particularly diffusion models, to predict dynamic crack evolution in realistic, complex semiconductor materials like AlN at the atomic scale, incorporating crucial microstructural features (e.g., void orientation, size distribution) under mechanical loading, has not been explored.

Therefore, this work aims to develop a novel diffusion-based deep learning framework specifically designed to predict the dynamic evolution of cracking in AlN at the atomic scale. Our model will explicitly incorporate material morphology, such as void characteristics, to enable high-fidelity predictions of crack initiation and propagation under mechanical load, offering a computationally efficient alternative to expensive MD simulations for this critical material failure behavior.

## 2. Methodology

### 2.1 Molecular Dynamics (MD) Simulations

Molecular Dynamics simulations were performed to model atomic-scale crack propagation in AlN using the Vashishta interatomic potential [36]. This potential incorporates two-body and three-body terms to accurately describe the mechanical and thermal behavior of crystalline and amorphous AlN. Initial atomic configurations consisted of a 3D thin-film structure measuring 623 Å × 622 Å × 22 Å. Each model contained a single elliptically shaped crack with randomized orientation, size, and position, as illustrated in Figure 1a. The semi-major axis length of each crack was uniformly sampled between 10–80 Å, while the aspect ratio (semi-minor to semi-major axis) followed a Gaussian distribution with a mean of 0.1 and standard deviation of 0.2. The total number of atoms for each model is around 800,000. A total of 1,000 distinct models were generated to ensure diversity in crack geometries.

The simulation protocol comprised two phases under an NPT ensemble at 300 K with zero pressure. First, energy minimization (relaxation) was performed. Subsequently, uniaxial tensile loading was applied along the horizontal axis at a strain rate of $5\times10^8$ s$^{-1}$ until reaching 5% total strain, with displacement controlled at the left and right boundaries. Periodic boundary conditions were employed to approximate bulk material behavior within finite-sized models. Simulations were executed using LAMMPS [37]. Atomic configurations were output at 10 non-uniformly spaced simulation steps (0, 50,000, 60,000, 70,000, 75,000, 80,000, 85,000, 90,000, 95,000, and 100,000 steps) to capture critical stages of dynamic crack evolution, particularly during later loading where changes accelerate. This sampling strategy captured the transformation from initial structures to deformed states exhibiting crack propagation (Figure 1), with evolution heavily influenced by initial crack orientation, size, and position.

Each atomic configuration (initial, deformed, and intermediate frames) was converted into a grayscale image of 192×192 pixels, representing atomic positions. This conversion resulted in a comprehensive dataset of 11,000 images (1,000 models × 11 frames per model). These images explicitly encode the microstructure evolution, including crack growth and orientation effects, serving as the foundational input data for training and validating the machine learning model.

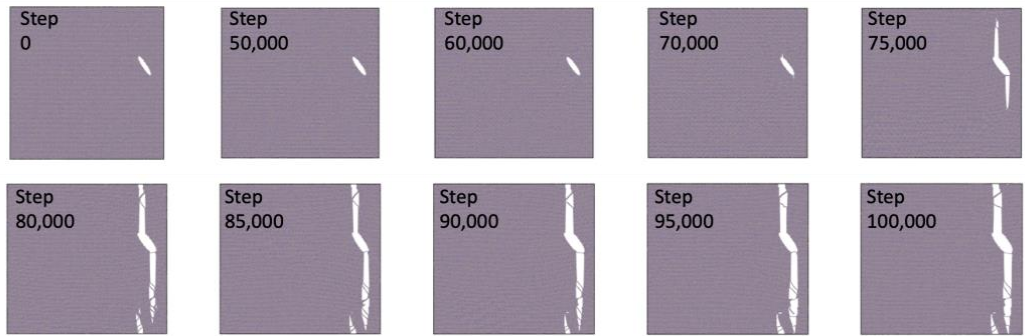

Figure 1: Atomic configurations from molecular dynamics (MD) simulations showing crack propagation evolution in AlN over 10 sequential time steps. Each frame captures the progressive fracture mechanism under tensile loading in the horizontal direction. These simulations serve as ground truth data for training the diffusion model.

## 2.2 Development of Diffusion-based Machine Learning Model

We developed a diffusion-based machine learning model building on prior work [33, 38-40] to predict crack evolution in AlN under mechanical loading, using atomic configuration images as the primary data representation. Our implementation adapts a progressive transformer diffusion architecture inspired by prior work [33], employing a U-Net backbone that outputs consecutive temporal frames capturing crack nucleation and propagation dynamics. Crucially, instead of text conditioning, we condition the model using microstructure embeddings—binary representations where material appears as black and cracks as white—fed through cross-attention layers. This approach steers the denoising process according to defect geometry while maintaining temporal consistency through 3D convolutional blocks and recurrent connections that correlate crack-tip motion between frames. The model jointly predicts all frames in a single diffusion pass using just 10 denoising steps, significantly enhancing computational efficiency compared to per-frame generation. Diffusion models generate data through a probabilistic process involving two phases: a forward process that systematically adds Gaussian noise to input images across multiple steps, and a learned reverse process where a neural network progressively denoises random noise to reconstruct realistic outputs. The model is trained to predict noise at intermediate steps, enabling accurate reconstruction of complex distributions. Key advantages include exceptional sample quality through iterative refinement, stable training dynamics, and flexible conditioning capabilities that incorporate material microstructure information.

Figure 2 illustrates the end-to-end workflow of our microstructure-conditioned diffusion model for predicting dynamic fracture evolution. During training (top), ground truth images—generated from molecular dynamics simulations of crack propagation—undergo a forward diffusion process where Gaussian noise is incrementally added over 10 steps. Simultaneously, the input microstructure (representing initial crack) is encoded into an embedding tensor. A U-Net (41.5M parameters) then learns to predict the noise at each timestep, conditioned on both the microstructure embedding and current noise level, using a specialized L2 loss against the actual noise. For inference (bottom), the trained model reverses this process: starting from pure Gaussian noise, it iteratively denoises the input over 10 reverse diffusion steps while guided solely by new microstructure embeddings, ultimately generating 10 high-fidelity frames that capture crack nucleation, propagation, and final fracture patterns without requiring atomistic simulations.

A key innovation of our approach is its exclusive reliance on crack images for training, eliminating the need for supplementary atomic-scale data such as potential energy distributions [33] or stress fields [41] that require expensive simulations. This simplification substantially reduces training costs while increasing practical utility, as crack images can be obtained directly from experimental microscopy data. The U-Net architecture contains 41,507,444 trainable parameters (detailed in [33, 38-40]), optimized using Adam with an initial learning rate of 0.0001 and batch size of 2. Training employed a specialized L2 noise-prediction loss across all frames, enabling the model to learn fracture physics implicitly from atomic simulation data.

The dataset comprised 1,000 samples featuring single-crack configurations, each providing an initial image plus 10 loading frames (11,000 total images). We allocated 90% for training and 10% for testing. To evaluate generalizability, we created an additional set of samples containing multiple cracks that were excluded from training, providing a rigorous out-of-distribution test for the model's predictive capabilities on more complex fracture scenarios.

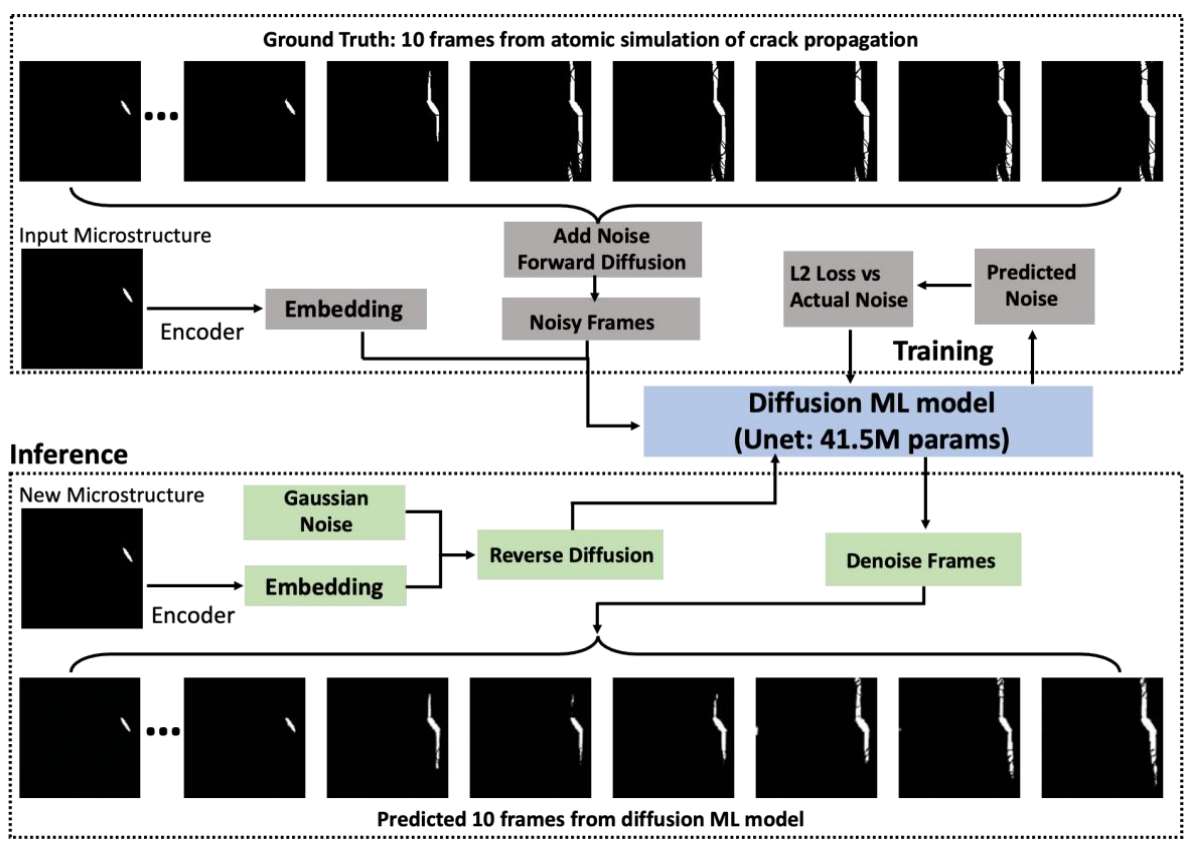

Figure 2: Training and inference workflow of the diffusion model for dynamic fracture prediction. During training (top), ground truth frames from atomic simulations undergo forward diffusion (noise addition), while a U-Net (41.5M parameters) learns to predict noise from microstructure-conditioned inputs. During inference (bottom), the trained model reverses diffusion to generate 10 fracture progression frames from Gaussian noise, guided solely by new microstructure embeddings.

## 3. Results and Discussion

### 3.1 Effects of Training Sample Size on Crack Propagation Prediction

A fundamental challenge in developing ML models for engineering applications is data scarcity, particularly when compared to the vast text corpora available for large language models. While atomistic simulations provide a more accessible data source than physical experiments, generating comprehensive datasets for complex phenomena like dynamic fracture remains computationally prohibitive. To balance predictive accuracy with practical feasibility, we focus on a simplified system: a single elliptical crack under uniaxial tension ($\leqslant 5\%$ strain). This constrained problem enables effective sampling of fracture physics while maintaining manageable computational costs.

To quantify sample size effects, we trained three diffusion models using 100, 500, and 1,000 atomistic simulation samples. Figure 3 reveals distinct optimization behaviors: the 100-sample model plateaus at significantly higher loss due to insufficient fracture mode diversity, while both 500- and 1,000-sample models converge to similar loss values after 200 epochs. This indicates diminishing effects beyond 500 samples for this simplified fracture system.

Prediction accuracy was further evaluated against MD ground truth (Figure 4). The 100-sample model exhibits substantial path deviation, failing to capture crack branches visible in simulations. In contrast, both the 500-sample and 1000-sample models achieve qualitative alignment with primary fracture paths, though the 1000-sample model demonstrates superior accuracy in predicting crack branching initiation at the applied strain of 0.04. This comparison confirms that sample size significantly governs prediction accuracy, but its influence decreases beyond 500 training samples in our system.

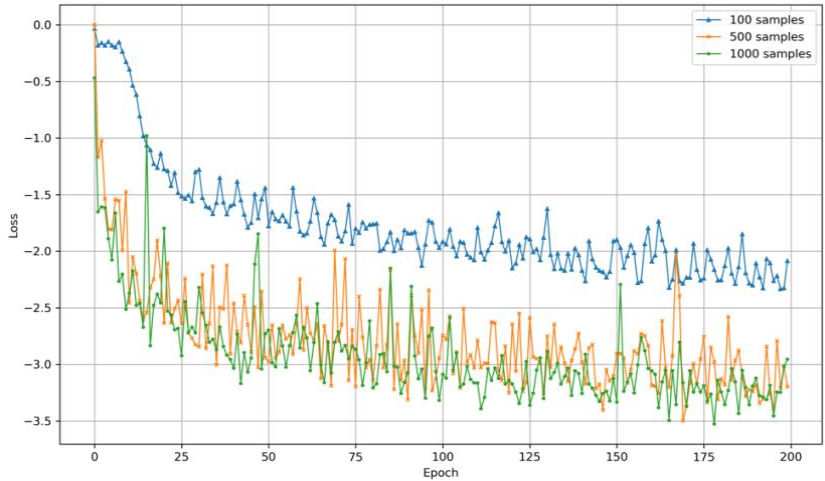

Figure 3: Log-scaled training loss evolution showing dataset size dependence. Models trained with 100, 500, and 1,000 atomistic simulation samples exhibit distinct convergence behaviors: the 100-sample model plateaus at higher loss values, while 500- and 1,000-sample configurations converge to similar loss after 200 epochs.

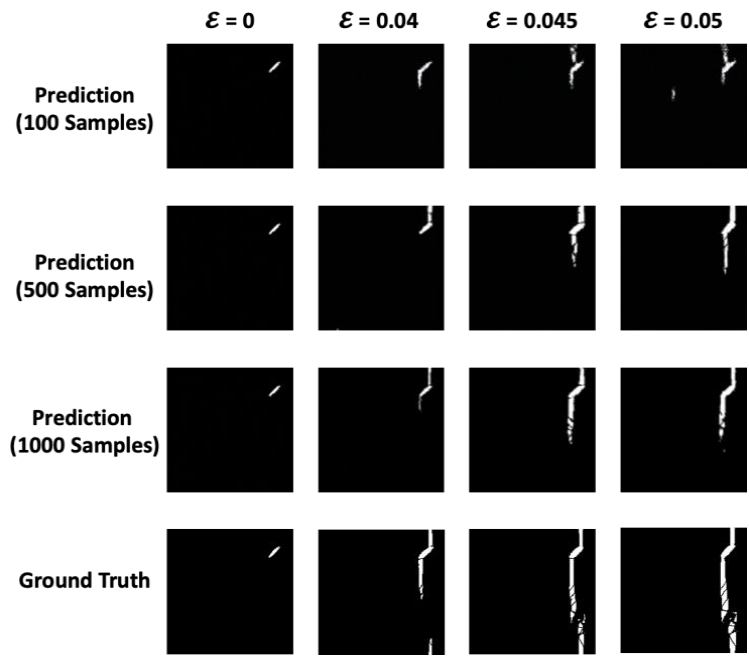

Figure 4: Crack propagation predictions versus MD ground truth for diffusion models trained with 100, 500, and 1,000 atomistic simulations. The 100-sample model shows significant path deviation, while the 500-sample model and 1000-sample models capture major crack branches observed in ground truth simulations.

## 3.2 Effects of Periodic Boundary Conditions on Crack Propagation Prediction

Atomistic simulations typically employ periodic boundary conditions (PBCs) to approximate bulk material behavior within finite-sized models. However, PBCs inherently influence the evolution of defects like cracks. A specific artifact arises when a propagating crack reaches a simulation box boundary: due to periodicity, the crack effectively "exits" one boundary and simultaneously "enters" the opposite boundary. This section investigates the impact of crack propagation across PBCs on machine learning predictions of crack paths, a phenomenon not explicitly addressed in prior studies [33-35, 41].

We compare ML predictions against ground truth simulations for two representative samples (Figure 5). For Sample A, where crack propagation occurs entirely within the simulation domain, the trained diffusion model successfully predicts the major crack branches observed in the ground truth. Conversely, for Sample B, when the primary crack crosses the top periodic boundary and re-enters through the bottom boundary at an applied strain of 0.04, the diffusion model fails to predict the subsequent branch initiating from this re-entry point.

This comparison confirms the trained ML model's proficiency in predicting physically realistic crack branching prior to PBC interactions. While the model correctly does not replicate artificial branching at periodic boundary re-entry points, a known simulation artifact, its predictive accuracy remains robust for fracture phenomena governed by intrinsic material behavior.

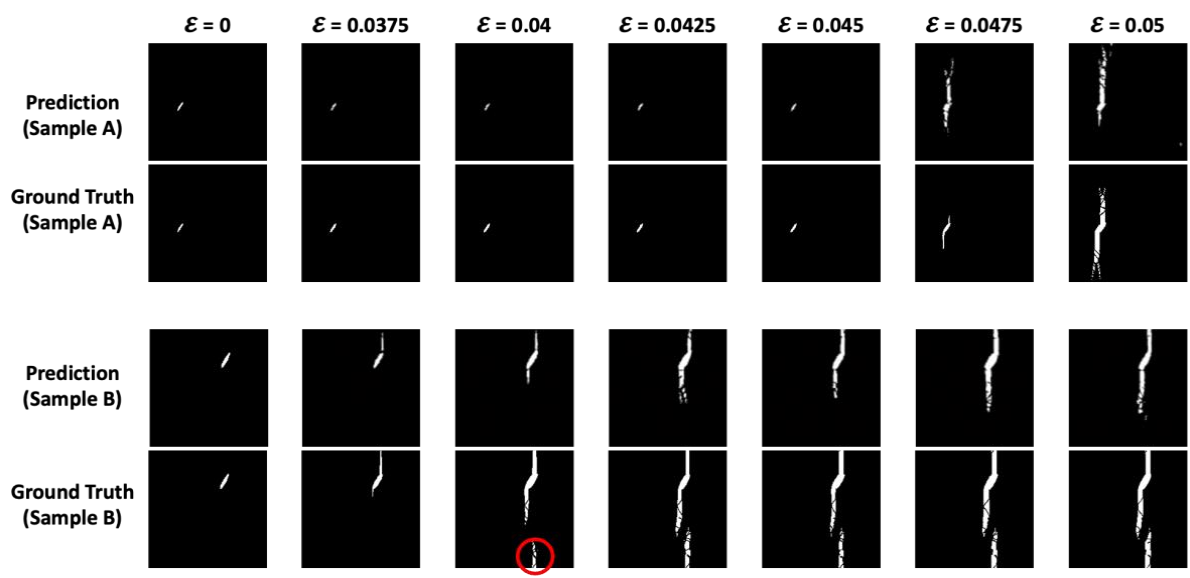

Figure 5: Effect of periodic boundary condition on crack propagation predictions. Sample A: Diffusion model prediction captures major crack branches versus MD ground truth. Sample B: At the applied strain of 0.04, crack propagation across periodic boundaries (circle) induces prediction failure, where the diffusion model misses secondary branching at the re-entry point (circle), demonstrating limitations in handling discontinuities induced by periodic boundary conditions.

## 3.3 Comparison of Fracture Details between Simulation and ML predictions

We compare atomic-scale fracture features from MD simulations with predictions by our trained diffusion model to evaluate crack propagation accuracy (Figure 6). Figure 6a displays the initial pre-cracked atomic configuration, while Figure 6b shows the fractured MD configuration at 0.04 tensile strain. Figure 6c provides a binary representation of this fractured state, and Figure 6d presents the diffusion model's prediction at equivalent strain. The diffusion model successfully reproduces major crack branches observed in MD results (Figures 6c–d), confirming its capability in capturing primary fracture pathways.

Detailed analysis reveals two complex phenomena: atomic-scale bridging ligaments between fracture surfaces (highlighted by yellow dotted circles), and crack propagation across the top periodic boundary with re-entry through the bottom boundary (marked by a red solid circle). For bridging ligaments, the diffusion model accurately predicts these connective features at annotated locations (Figure 6d), demonstrating its capacity to resolve essential physical fracture mechanisms. Conversely, the model fails to predict the artificial crack branch at the periodic boundary re-entry site (red circle, Figure 6d), consistent with Section 3.2's findings regarding boundary-induced artifacts.

The trained diffusion model thus reliably predicts physically meaningful fracture characteristics, including major crack branches and atomic-scale bridging ligaments, validating its ability to capture intrinsic material fracture behavior. However, it cannot replicate unphysical artifacts stemming from periodic boundary constraints, highlighting a domain limitation for scenarios involving boundary-transcending crack propagation. This distinction underscores the model's accuracy to physical fracture behaviors while exposing boundaries for its predictive applicability.

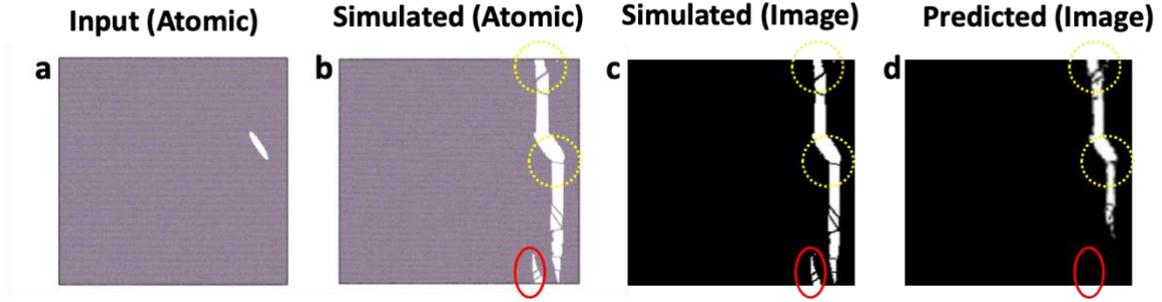

Figure 6: Crack propagation prediction by the trained diffusion model against MD ground truth. The model successfully predicts major crack branches and atomic-scale crack bridging ligaments (yellow dotted circles) but fails to capture the artificial crack branch at the periodic boundary re-entry point (red solid circle).

## 3.4 Effects of Initial Crack Geometry on Propagation Behavior

We investigate how initial crack geometry influences propagation by comparing MD simulations with diffusion model predictions across six distinct samples (top row, Figure 7). These samples vary in crack size, orientation, shape, and location within the simulation domain. Crucially, crack orientation relative to loading direction determines propagation initiation: Horizontally aligned cracks (e.g., Sample 4) parallel to the tensile loading direction remain stable, while inclined cracks propagate due to resolved shear stress components. Our diffusion model accurately predicts this directional dependence, correctly identifying non-propagation in Sample 4 and propagation in all other orientations (middle row, Figure 7).

For propagating cracks, we observe consistent bifurcation patterns where cracks initiate simultaneously from both tips of the initial defect (bottom row, Figure 7). The resulting branches propagate vertically, orthogonal to the horizontal tensile loading direction, demonstrating crack path sensitivity to far-field stress states. The diffusion model successfully reproduces this dual-tip initiation and vertical branching behavior across all relevant samples, confirming its ability to capture orientation-dependent propagation physics.

Consistent with Sections 3.3, we observe two similar phenomena in propagating samples: atomic-scale bridging ligaments between fracture surfaces and periodic boundary artifacts where cracks cross the top boundary and re-enter through the bottom. The diffusion model reliably predicts ligament formation but fails to capture boundary-induced artificial branching at re-entry points. This validation across geometrically diverse samples confirms the robustness of Section 3.3's findings regarding the model's capacity to replicate physical mechanisms while remaining insensitive to simulation artifacts.

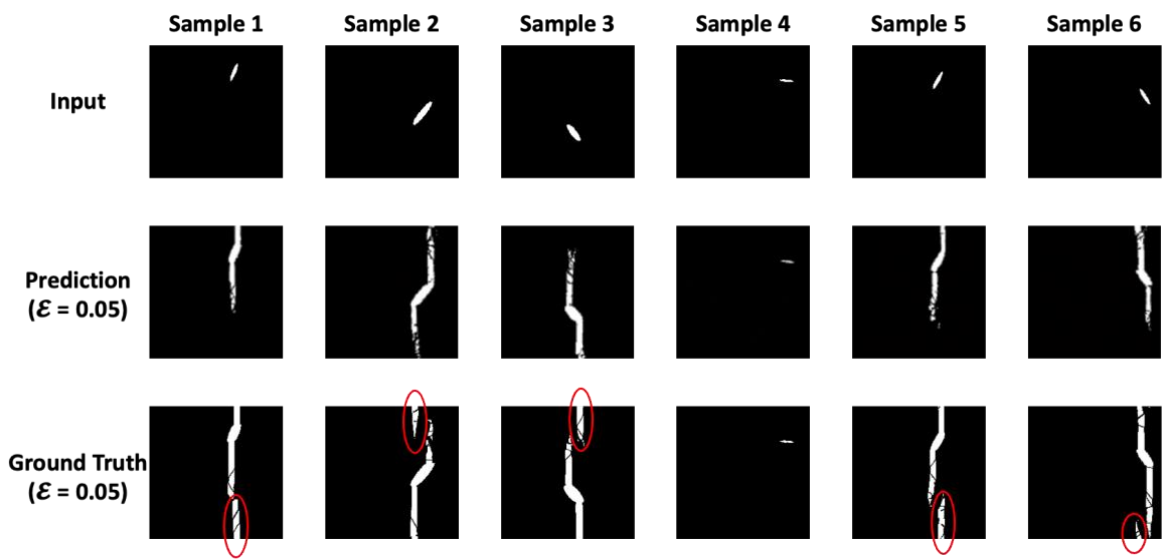

Figure 7: Crack propagation predictions by the diffusion model versus MD ground truth for samples with varying initial crack geometries. Top row: Initial configurations showing crack geometry variations. Middle row: Diffusion model predictions at 0.05 applied strain. Bottom row: Corresponding MD simulation results at 0.05 strain. Red solid circles indicate artificial crack branches at periodic boundary re-entry points.

**3.5 Generalizability of Diffusion ML Models for Complex Crack Configurations**

To explore the predictive boundaries of diffusion models beyond their training domain, we evaluate samples containing multiple initial cracks, configurations absent from the training data (single-crack specimens). Figure 8 presents three samples (N1–N3) with 4–5 stochastically distributed cracks. In Sample N1, MD simulations reveal propagation initiates at the left crack rather than the largest defect, contrary to conventional expectations. This occurs because the largest crack aligns near-horizontally with the tensile axis, minimizing resolved shear stress, while the left crack's inclined orientation maximizes driving forces. The diffusion model correctly predicts this non-intuitive initiation site, demonstrating its capacity to resolve stress-driven propagation physics beyond geometric size dependencies.

For Sample N2, propagation initiates at the largest crack due to its favorable inclination relative to the tensile direction. The diffusion model accurately predicts both initiation and precise branching locations along this crack's surface. In Sample N3, the model successfully forecasts propagation from the largest defect and its eventual connection to a neighboring crack in the upper-right region. A minor discrepancy arises where the model omits a top-boundary crack branch, an artifact of periodic boundary conditions consistent with Sections 3.2. These results collectively validate the model's ability to handle multi-crack interactions despite training exclusively on single-crack specimens.

We further test generalizability on a geometrically complex sample with 10 initial cracks (Figure 9). MD simulations reveal two dominant propagation paths: a left-path connecting three vertically aligned cracks and a right-path that crosses the top boundary and re-enters through the bottom (a PBC artifact). The diffusion model accurately predicts the left-path and the right-path's initiation at 0.04 strain. However, when the right-path approaches the top boundary, the diffusion model, insensitive to PBC artifacts, diverges by predicting a physically plausible third path from a similarly inclined crack instead of simulating re-entry. This behavior indicates the model prioritizes material-intrinsic physics over PBC artifacts.

Crack propagation is a dynamic process, which inherently is a time-series problem. Crucially, diffusion models embed temporal information directly within their architecture through learned timestep encodings. This design fundamentally prevents error accumulation across sequential predictions, a limitation observed in recurrent approaches like Long Short-Term Memory (LSTM) networks where uncertainties propagate between timesteps. The temporal stability inherent to diffusion modeling significantly enhances prediction accuracy for crack evolution trajectories.

Collectively, these results demonstrate the diffusion model's robust generalizability to multi-crack systems geometrically distinct from its training data. While PBC-induced artifacts remain beyond its predictive scope, the model consistently captures stress-driven initiation, crack coalescence, and branching phenomena across complex configurations. This underscores its value for simulating realistic fracture behavior where defect interactions govern failure.

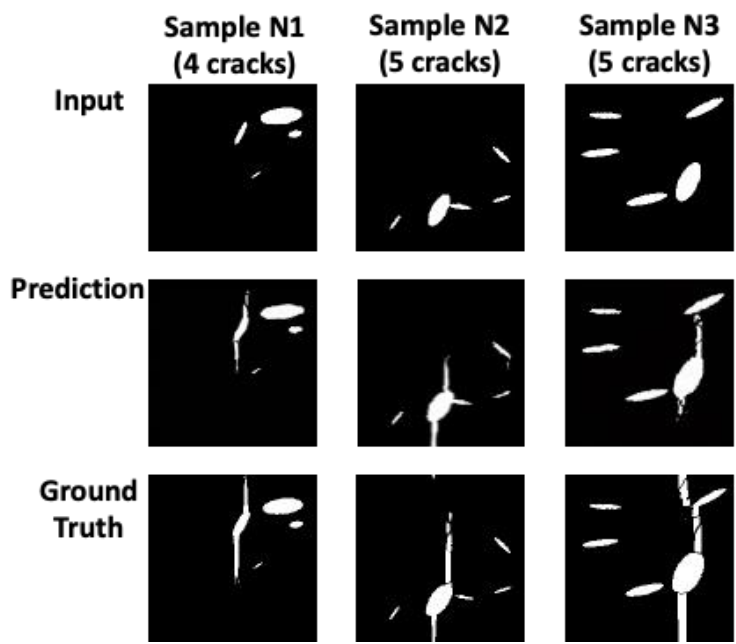

Figure 8: Generalizability assessment for multi-crack systems (4-5 initial cracks) versus MD ground truth. Top row: Initial configurations showing stochastically distributed cracks. Middle row: Diffusion model predictions. Bottom row: Corresponding MD simulation results.

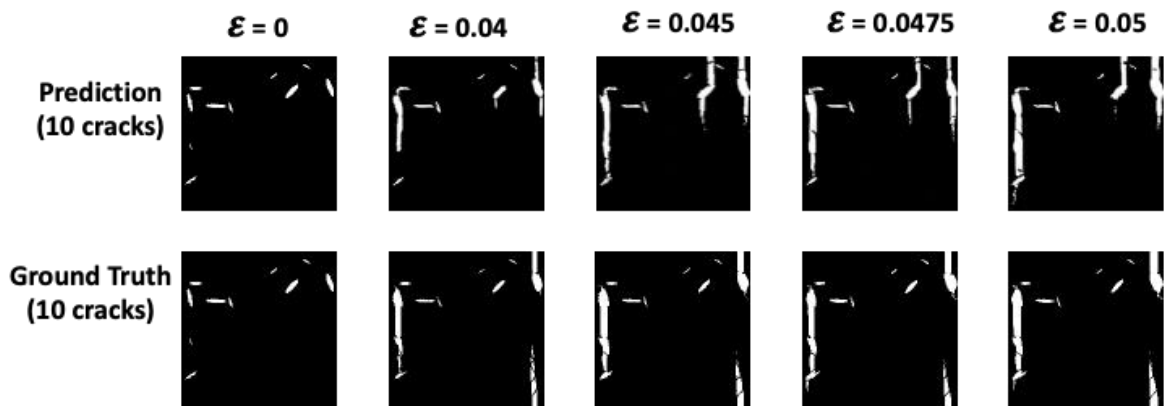

Figure 9: Predictive capability assessment for complex multi-crack system (10 initial cracks) versus MD ground truth. Top row: Diffusion model predictions at progressive strains. Bottom row: Corresponding MD simulation results.

## 4. Conclusion

This study develops a diffusion-based generative model for predicting atomic-scale crack evolution in AlN under mechanical loading. By conditioning the model on microstructure embeddings derived solely from initial crack geometries, our model accurately forecasts dynamic fracture processes, including crack branching, ligament formation, and path progression, while achieving a significant speedup over conventional molecular dynamics simulations. The model demonstrates exceptional accuracy in capturing stress-driven crack initiation patterns and defect interactions across diverse configurations, validating its capacity to learn complex fracture physics directly from atomic simulation data without requiring supplementary stress or potential energy fields.

The diffusion model demonstrates two defining boundaries in predictive capability: First, it reliably reproduces physical fracture mechanisms, including atomic-scale bridging ligaments and stress-driven branching, while disregarding artifacts induced by periodic boundary conditions, reflecting its inherent bias toward material-intrinsic behavior. Second, though generalizing effectively to multi-crack systems beyond its training domain, its performance remains constrained when cracks cross simulation boundaries, violating the model's physics-consistent generation paradigm. These limitations reveal a critical strength: the model's capacity to discriminate physical phenomena from numerical artifacts establishes its value for real-world material failure prediction. However, they also highlight the need for simulation-aware training strategies when periodic boundary effects dominate, defining clear operational boundaries for predictive applicability.

Our microstructure-conditioned diffusion framework bridges atomic-scale accuracy with computational feasibility, enabling rapid exploration of fracture mechanics in semiconductor materials. By eliminating dependency on high-cost MD simulations for crack prediction, this approach opens avenues for optimizing AlN thin-film fabrication and failure-resistant device design. Future work will extend this paradigm to temperature-dependent cracking and multi-material systems while integrating experimental microscopy data for direct experimental validation, ultimately advancing predictive capabilities for next-generation semiconductor reliability.

## Acknowledgements

This work is supported by the National Science Foundation under Award Number CMMI-2436919.

## Conflict of Interest

The authors have no conflicts to disclose.

## Data Availability

The data will be provided by the authors based on request.